\documentclass[a4paper]{ESASPCS13Style}
\usepackage{epsfig}

\begin{document}

\title{Dynamo action in M-dwarfs}

\author{S. B. F. Dorch\inst{1}, B. V. Gudiksen\inst{2}, and H.-G. Ludwig\inst{3}}
  \institute{The Niels Bohr Institute for Astronomy, Physics and Geophysics,
     Juliane Maries Vej 30, DK-2100 Copenhagen {\O}, Denmark
  \and Institute of Theoretical Astrophysics, P.O. Box 1029 Blindern,
     N-0315 Oslo, Norway
  \and Lund Observatory, Box 43, SE-221 00 Lund, Sweden}

\maketitle

\begin{abstract}

Magnetic activity in M-dwarfs present enigmatic questions: On the
one hand they have higher field strengths and larger filling
factors than the magnetic field on the Sun, on the other hand,
they are fully convective and their atmospheres are more neutral,
hence they do not have an undershoot layer for magnetic flux
storage and as we show here, cannot have small-scale dynamo action
in their photospheres either. We present a discussion of these
facts and propose a new numerical model to investigate M-dwarf
magnetism.


\end{abstract}

\section{Introduction}

M-type dwarf stars display a high degree of magnetic activity.
Both indirect tracers as X-ray and $\mathrm{H}_\alpha$ emission,
as well as Zeeman broadening of magnetically sensitive
photospheric lines has been observed (Johns-Krull \&\ Valenti
1996, Kochukhov et al. 2001). The photospheric lines show no
rotational modulation or net polarization, indicating that fields
of small scale relative to the stellar radius exist on the
surfaces of M-dwarfs. These fields can have strengths of a few kG
and can cover a substantial area with filling factors up to 50\%.

On the theoretical side of things, it has long been considered a
problem that M-dwarfs are fully convective, lending no helping
hand to the storage of magnetic flux through the presence of a
stable stratified convective undershoot layer (e.g.\ van
Ballegooijen \cite{Ballegooijen1982}). Furthermore, as we show
below (see also Dorch \& Ludwig \cite{Dorch+Ludwig2002}), it seems
that their upper atmospheres are not capable of providing a
small-scale network magnetic field, similar to that in the Sun,
sometimes assumed to be stemming from local small-scale dynamo
action in the photosphere (Cattaneo \cite{Cattaneo1999}). However,
recent global scale simulations of M-dwarfs seem to render the
storage problem obsolete (Dobler \cite{Dobler2004}), and in this
contribution we propose future possibilities regarding numerical
simulations of M-dwarf magnetic activity. First, however, we
discuss results from models of M-dwarf local dynamo action.

\section{Local small-scale dynamo action}

\begin{figure}
\resizebox{\hsize}{!} {\includegraphics[]{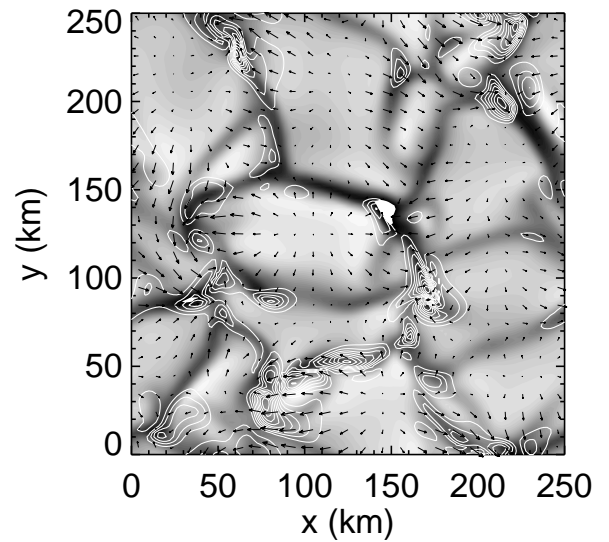}}
\caption{Snapshot of the vertical component of the magnetic field
at optical depth unity (contours), emergent intensity (grey
tones), and the velocity field (arrows) towards the end of an
evolutionary sequence of a model with Rm=20.} \label{fig1}
\end{figure}

In order to study local small-scale dynamo action in the
photosphere of M-dwarfs we adopted a velocity field from a
radiation-hydrodynamics simulation of a prototypical M-dwarf
atmosphere from Ludwig et al.\ (2002), see Fig.\ \ref{fig1}: Here
$T_{\rm eff}=2800$ K, $\log g=5.0$ and the chemical composition is
solar. Magnetic fields can potentially influence the flow
structure significantly provided the field is strong. However,
this corresponds to the full MHD case which is beyond the scope of
this contribution. We selected a sequence which comprises roughly
10 convective turn-over time scales. The sequence consists of 150
snapshots of the flow field, each comprising \mbox{$125\times 125
\times 82$} grid points corresponding to \mbox{$250\times
250\times 87$ km$^3$}. At any instant in time about 10 granular
cells were present in the computational domain, ensuring a
statistically representative ensemble.

To compute the evolution of the magnetic field in responds to the
adopted velocity field, we assume the kinematic regime of MHD,
where one can neglect the back-reaction of the magnetic field on
the fluid motions. Then solving the MHD equations reduces to the
problem of seeking the solution to the time-dependent induction
equation:
\begin{equation}
 \frac{\partial {\bf B}}{\partial t} = \nabla\times ({\bf u}\times {\bf B})
 + \eta \nabla^2 {\bf B}, \label{induction.eq}
\end{equation}
where ${\bf u}$ is the prescribed velocity field, ${\bf B}$ the
magnetic field, and $\eta$ the magnetic diffusivity that. We
assume a spatially constant magnetic diffusivity that we vary by
setting the magnetic Reynolds number Rm defined as $\mathrm{Rm} =
{\rm U} \ell/\eta$ (where U is a characteristic velocity, $\ell$ a
characteristic length scale, and $\eta$ the magnetic diffusivity).

We solve Eq.\ (\ref{induction.eq}) using staggered variables on
the grid of the hydrodynamical flow field. The numerical method
was originally developed by Galsgaard and others (Galsgaard \&
Nordlund \cite{Galsgaard+Nordlund97}) for general MHD purposes,
but a special version is the code used by Archontis, Dorch, \&
Nordlund (\cite{Archontis+ea03}) for studying dynamo action in
prescribed flows. The kinematic approximation to the MHD equations
becomes inaccurate when the magnetic field becomes sufficiently
strong, limiting us to weak fields.

\begin{figure}[t!]
\resizebox{\hsize}{!} {\includegraphics[]{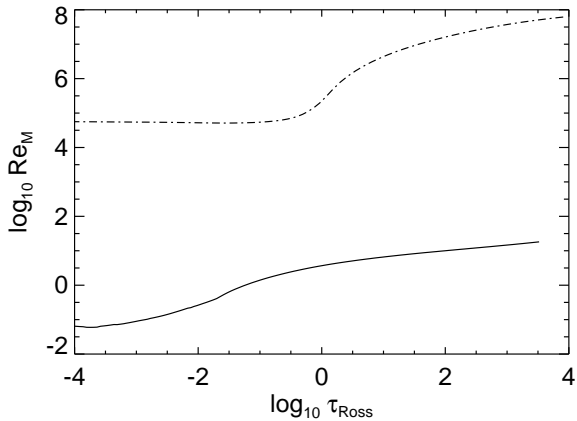}}
\caption[]{The magnetic Reynolds number as a function of Rosseland
optical depth in a \mbox{$T_{\rm eff}=2800 {\rm K}$} M-dwarf
(solid line) and a solar model atmosphere (dashed line). Note the
low magnetic Reynolds number in the M-dwarf model, primarily
reflecting the low electric conductivity of the stellar gas in the
rather cool M-dwarf atmosphere.} \label{fig2}
\end{figure}

In the following we discuss kinematic MHD models with Rm as low as
20. In the astrophysical context, low values of Rm are uncommon,
mostly due to the large spatial scales usually involved. However,
in M-dwarf atmospheres we can be confronted with the situation of
rather low Rm (Meyer \&\ Meyer-Hofmeister 1999): Fig.\ \ref{fig2}
shows  Rm in the M-dwarf model ($\ell=80~ {\rm km}$, $v=0.16~ {\rm
km/s}$) as well as the Sun ($\ell=1500~ {\rm km}$, $v=2.4~ {\rm
km/s}$). Rm primarily reflects the run of the electric
conductivity in the atmosphere, which in turn is mostly controlled
by the electron to gas pressure. The conductivity has been
evaluated assuming a weakly ionized plasma.

At sufficiently cool temperatures Rm reaches order unity in the
surface layers. This is a consequence of the declining electron
density, the shrinking of spatial scales, and smaller convective
velocities: This refers to the surface layers, but qualitatively
we expect a strong increase of Rm with depth, and beyond a certain
depth the regime of ${\rm Rm}>1$ is reached again.  However, gas
motions in this depth will generally be slower and the tangling of
magnetic field lines less rapid, which may reduce the efficiency
of chromospheric and coronal heating. Whether this plays a
r\^{o}le for the observed decline of stellar activity at the
transition from M- to L-dwarfs (Gizis et al. 2000) is presently a
matter of debate (Mohanty et al. 2002, Berger 2002).

A flow is a fast kinematic dynamo when the exponential growth rate
$\gamma$ is positive. That is, fast dynamo action requires a
continuous increase of magnetic energy, even in the limit of
vanishing diffusivity. This limit is relevant because most
astrophysical systems have ${\rm Rm} \gg 1$ and small but non-zero
$\eta$. It is believed that turbulent astrophysical systems are
fast dynamos operating at ${\rm Rm} \gg 1$. When Rm increases, the
length scale of magnetic islands decreases and scales as ${\rm
Rm}^{-\frac{1}{2}}$. There is a maximum value of {\rm Rm}, which
can be achieved with our numerical resolution $\Delta x = 2$ km,
and therefore the largest magnetic Reynolds number that we can
allow is of the order of 400 corresponding to the Nyquist
wavelength $2~ \Delta x$. In the following, we concentrate on
situations with ${\rm Rm} \le 400$.

In the absence of non-linear effects, in case of a dynamo, one
expects a continuing exponential growth of magnetic energy ${\rm
E}_{\rm M}$. Figure \ref{fig3} shows the result in terms of ${\rm
E}_{\rm M}(t)$ for five different models corresponding to varying
{\rm Rm}:
\begin{figure}
\resizebox{\hsize}{!} {\includegraphics[]{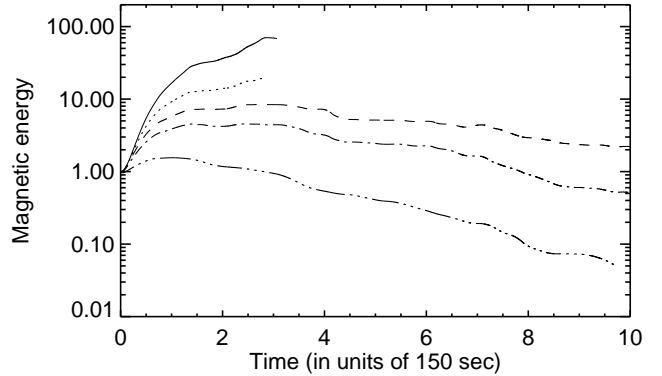}}
\caption{The total magnetic energy ${\rm E}_{\rm M}/{\rm E}_0$ as
a function of time in units of 150 seconds (the typical convective
turn-over time). Five models are presented, with different amounts
of diffusion: {\rm Rm} = 800 (full), 400 (dotted), 200 (dashed),
100 (dashed dotted), and 20 (dashed triple dotted curve).}
\label{fig3}
\end{figure}
Most of the models in fact are dominated by decaying modes, with
negative growth rates. In particular, the case with {\rm Rm} = 20
is clearly an example of an anti-dynamo: The diffusion works
faster than the flows can sweep up the field and concentrate it in
the inter-granular down-draft lanes, and the dominant magnetic
mode is a decaying one. Increasing Rm decreases $\gamma$, so that
for {\rm Rm} = 200 the decay time is 40 time longer than at {\rm
Rm} = 20. In terms of providing dynamo action, the most promising
cases are those with lower diffusion and {\rm Rm} $>$ 300--400: At
{\rm Rm} = 400 a growing mode seems to be dominating.

In the high diffusion case the magnetic field varies smoothly
across the domain, and its power peaks on scales approximately
$\ell \approx 50~ {\rm km}$ when considering $B_\mathrm{z}$ (see
Fig.\ \ref{fig1}). At higher Rm more small scales are generated
mostly around the down-draft lanes.

\section{Discussion and conclusion}

Little is known about the structure of magnetic fields in the
photospheres of M-dwarfs. From an observational point of view one
would like to get some input from theory that would alleviate the
problem of disentangling field strength and filling factor. To
this end we started to investigate the kinematic effect of the
convective velocity field on a magnetic seed field. This has
become possible due to recent progress in the hydrodynamical
modelling of atmospheres in the regime of cool M-dwarfs (Ludwig,
Allard, \&\ Hauschildt 2002).

\begin{figure}
\resizebox{\hsize}{!} {\includegraphics[]{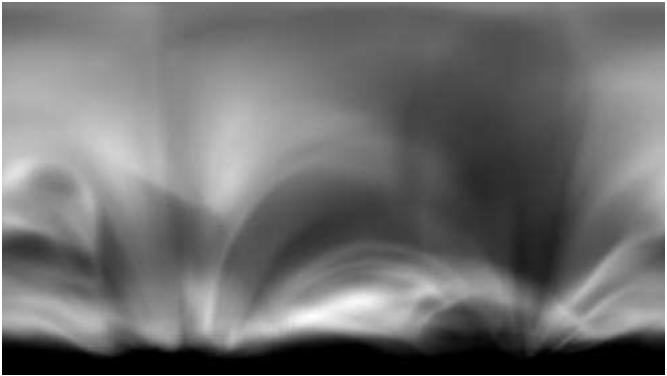}}
\caption{Illustration of loops formed in a solar corona simulation
using a synthetic TRACE 195 {\AA} filter.} \label{fig4}
\end{figure}

We performed kinematic studies of the evolution of small-scale
magnetic fields in the surface layers of M-dwarfs. We solved the
induction equation for a prescribed velocity field, magnetic
Reynolds number {\rm Rm}, and boundary conditions in a Cartesian
box, representing a volume comprising the optically thin stellar
atmosphere and the uppermost part of the optically thick
convective envelope. We find dynamo action for ${\rm Rm} \ge 400$
and growth time scales of the magnetic field are comparable to the
convective turn-over time scale ($\approx 150 {\rm sec}$). The
convective velocity field concentrates the magnetic field in
sheets and tubular structures in the inter-granular down-flows.
Perhaps surprisingly, Rm is of order unity in the surface layers
of cooler M-dwarfs, rendering the dynamo inoperative.  In all
studied cases we find a rather low spatial filling factor of the
magnetic field.

\begin{itemize}
\item
The geometry of small-scale magnetic fields looks similar to the
situation for the Sun. The basic reason is that M-dwarf
granulation is qualitatively similar to solar granulation.

\item
There are differences due to potentially quite different magnetic
diffusivities.

\item Depending on Rm, we get or do not get local dynamo action.
\end{itemize}

Future work along the lines of understanding M-dwarf magnetism
will involve a more complicated treatment than that discussed
above: It is our goal to combine both numerical theoretical
modelling and detailed observations to begin to understand how
M-dwarfs dynamos operate. We will use numerical models, such as
the detailed radiative hydrodynamic models used here, together
with detailed coronal models similar to those that have recently
been successfully applied to the Sun (Gudiksen \& Nordlund 2001).
By constructing a model that connects the surface and subsurface
magnetic field to the corona, it is our hope that the energetic
events observed in the coronae of M-dwarfs will provide a link to
understanding the dynamo. This goal will be achieved by employing
a sequence from a radiative hydrodynamic model as a lower velocity
boundary condition for a coronal model, yielding an output that
can be compared directly to observations. The unknown magnetic
surface topology enters as a ``free parameter" as an initial lower
boundary condition that is driven to reconnect by the braiding
surface motions.

\begin{acknowledgements}
This work was supported by the Danish Natural Science Research
Council.
\end{acknowledgements}

\end{document}